# Simulation of Un-Symmetrical 2-Phase Induction Motor

[1] S. F. Bandhekar, [2] S. N. Dhurvey, [3] P. P. Ashtankar

[1, 2, 3] Priyadarshini Institute of Engineering and Technology
Nagpur, Maharashtra, India

**Abstract**
The equations of unsymmetrical 2-phase induction motors are established and a computer representation is developed from these equations. Computer representation of single phase motors are developed by extension and modification of the unsymmetrical 2-phase induction motors representation. These equations of an unsymmetrical 2-phase induction motors are describe the dynamic performance of equations of unsymmetrical 2-phase induction motors. The system is simulated to verify its capability such as input phase voltage, stator and rotor currents, electromagnetic torque and rotor speed. The performance of an unsymmetrical 2-phase induction motors is simulated using MATLAB simulink.

## 1. Introduction

The unsymmetrical 2-phase induction motors are generally used in single phase applications such as refrigerators, washing machine, clothes dryers, furnace fans, garbage disposals and air conditioners. In case of the unsymmetrical 2-phase induction motors a starting torque is produced by designing the stator windings to be electrically different and in some cases, placing a capacitor in series with one of the windings. The reference frame theory is used to establish the voltage equations in the stationary reference frame which are valid for transient and steady state condition.

The phasor equations valid for predicting steady state operation are established from these equations for stator voltages of any periodic waveform. The standard steady state equivalent circuits are then derived from these equations using symmetrical component theory. The theory of operation of the unsymmetrical 2-phase induction motors is applicable to a wide variety of single phase induction motors.

Therefore the equivalent circuits and the computer representation of the unsymmetrical 2-phase induction motors is developed and then modified and extended to describe the dynamic performance of various types of single phase induction motors. A 2-phase motor with identical rotor windings and nonsymmetrical stator windings is considered an unsymmetrical 2-phase induction motor. In this analysis of this type of motor, it is generally assumed that

(1) Each stator windings is distributed to produce sinusoidal mmf wave in space.

(2) The rotor mmf waves can be considered as sinusoidal having the same number of poles as the corresponding stator mmf wave.

(3) The air gap is uniform.

(4) The magnetic circuit is linear.

The equations which describes the transient and steady state performance of an unsymmetrical 2-phase induction motors can be established by considering the elementary 2-pole motor shown in fig.1. 2-phase induction motor is composed of two asymmetrical windings.

Therefore the number of auxiliary winding usually has fewer turns than main winding and displaced 90 electrical degrees between the windings. Fig 1 shows the schematic view of a 2-phase induction motors.

The auxiliary winding (α) and main winding (β) are not identical sinusoidal distributed windings but are arranged in space quadrature. The stator windings are unsymmetrical; the windings have an unequal resistance and unequal number of turns. The rotor windings are identical; i.e. the windings have an identical resistance and identical number of turns.

The equivalent circuits represented the unsymmetrical 2-phase induction motors in stationary (αβ) reference frame are shown in fig 2.

The dynamic model equations of an unsymmetrical 2-phase induction motors can be written in αβ reference frame variables.

## 2. Tables, Figures and Equations

2.1 Figures

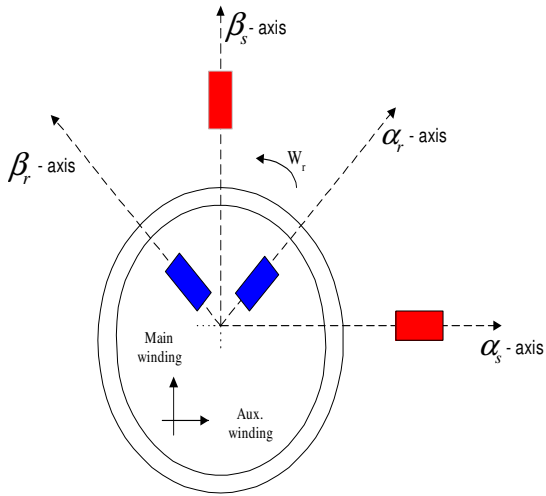

Fig. 1 Unsymmetrical 2-phase Induction Motor

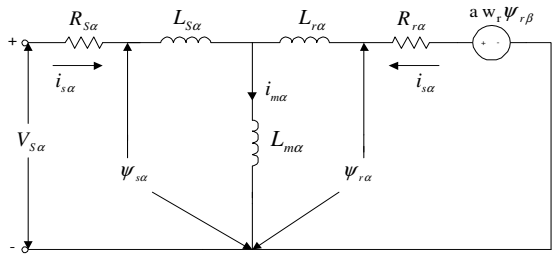

Fig.(a) :- Auxiliary winding in $\alpha$- axis

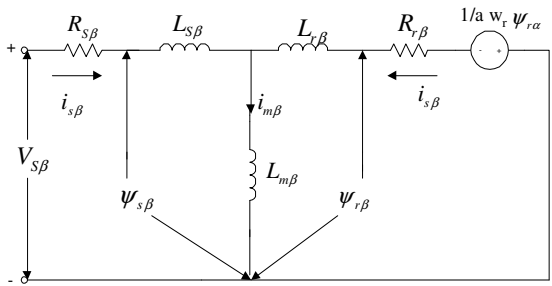

Fig.(b) :- Main winding in $\beta$ - axis

Fig. 2 Equivalent circuits of an unsymmetrical 2-phase Induction Motor in the stationary (α β) reference frame.

2.2 Equations

It is convenient to develop the computer representation of unsymmetrical 2-phase induction motors and then modified this representation to simulate various types of single phase applications.

The stator voltage equations are written as,

$$V_{s\alpha} = R_{s\alpha} i_{s\alpha} + \frac{d}{dt} \psi_{s\alpha} \quad (1)$$

$$V_{s\beta} = R_{s\beta} i_{s\beta} + \frac{d}{dt} \psi_{s\beta} \quad (2)$$

The rotor voltage equations are written as,

$$V_{r\alpha} = 0 = R_{r\alpha} i_{r\alpha} + \frac{d}{dt} \psi_{r\alpha} + a\,\omega_r \psi_{r\beta} \quad (3)$$

$$V_{r\beta} = 0 = R_{r\beta} i_{r\beta} + \frac{d}{dt} \psi_{r\beta} - \frac{1}{a}\omega_r \psi_{r\alpha} \quad (4)$$

The component of stator and rotor flux linkages can be expressed as:

$$\psi_{s\alpha} = L_{s\alpha} i_{s\alpha} + L_{m\alpha} i_{r\alpha} \quad (5)$$

$$\psi_{s\beta} = L_{s\beta} i_{s\beta} + L_{m\beta} i_{r\beta} \quad (6)$$

$$\psi_{r\alpha} = L_{m\alpha} i_{s\alpha} + L_{r\alpha} i_{r\alpha} \quad (7)$$

$$\psi_{r\beta} = L_{m\beta} i_{s\beta} + L_{r\beta} i_{r\beta} \quad (8)$$

Using equations (5)–(8), the stator and rotor currents equations are given by :

$$i_{s\alpha} = \frac{L_{r\alpha} \psi_{s\alpha} - L_{m\alpha} \psi_{r\alpha}}{L_{s\alpha} L_{r\alpha} - L_{m\alpha}^2} \quad (9)$$

$$i_{s\beta} = \frac{L_{r\beta} \psi_{s\beta} - L_{m\beta} \psi_{r\beta}}{L_{s\beta} L_{r\beta} - L_{m\beta}^2} \quad (10)$$

$$i_{r\alpha} = \frac{L_{s\alpha} \psi_{r\alpha} - L_{m\alpha} \psi_{s\alpha}}{L_{s\alpha} L_{r\alpha} - L_{m\alpha}^2} \quad (11)$$

$$i_{r\beta} = \frac{L_{s\beta} \psi_{r\beta} - L_{m\beta} \psi_{s\beta}}{L_{s\beta} L_{r\beta} - L_{m\beta}^2} \quad (12)$$

The equation of electromagnetic torque produced by the machine is then given by the equation :

$$T_e = p_p (L_{m\beta} i_{s\beta} i_{r\alpha} - L_{m\alpha} i_{s\alpha} i_{r\beta}) \quad (13)$$

The mechanical dynamic is modeled by the equation

$$J \frac{d}{dt} \omega_r = T_e - T_L \quad (14)$$

The Simulink model of an unsymmetrical 2-phase Induction Motor is created from above equations is shown in fig 3.

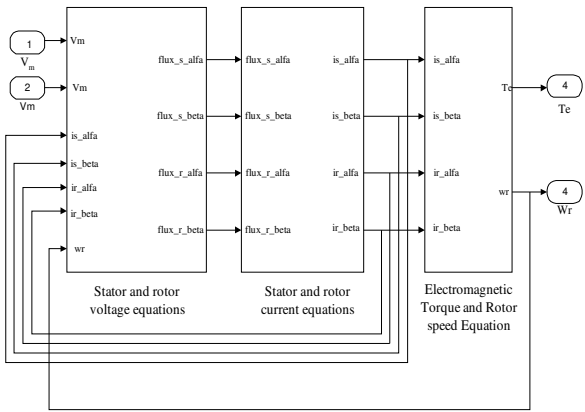

Fig. 3 Simulink Model of Unsymmetrical 2-phase Induction Motor

## 3. Simulation of Unsymmetrical 2-phase Induction Motor

The simulink model of unsymmetrical 2-phase induction motor which is shown in figure 5.4 is now developed and simulated in MATLAB/simulink environment. The motor parameters are tabulated in table 1 and the rated are 230V, 50 Hz, ¼ HP, and 4 pole. In all simulated cases, the load torque is fixed at 1.0096 Nm.

Table 1 :- 2-phase Induction Motor parameters

| $R_{s\alpha}$ = 7.14 Ω | $L_{s\alpha}$ = 0.2549 H | $L_{m\alpha}$ = 0.2464 H |
|---|---|---|
| $R_{s\beta}$ = 2.02 Ω | $L_{s\beta}$ = 0.1846 H | $L_{m\beta}$ = 0.1772 H |
| $R_{r\alpha}$ = 5.74 Ω | $L_{r\alpha}$ = 0.2542 H | $J = 2.92 \times 10^{-3}$ Kg-m$^2$ |
| $R_{r\beta}$ = 4.12 Ω | $L_{r\beta}$ = 0.1828 H | a = 1.18 |

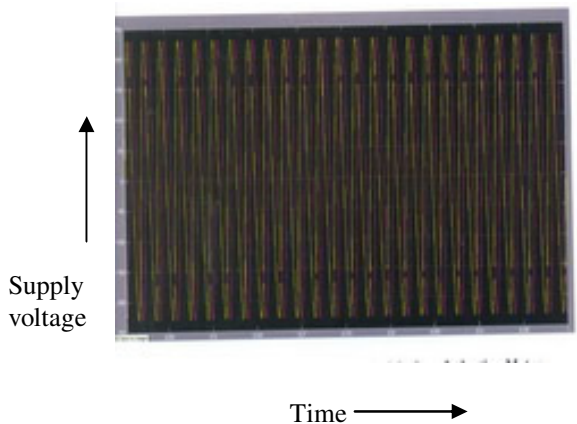

Fig.4 Supply voltage of Unsymmetrical 2-phase Induction Motor

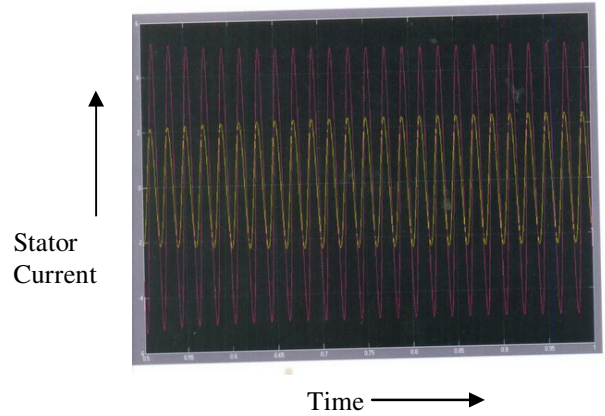

Fig. 5 Stator Current of Unsymmetrical 2-phase Induction Motor

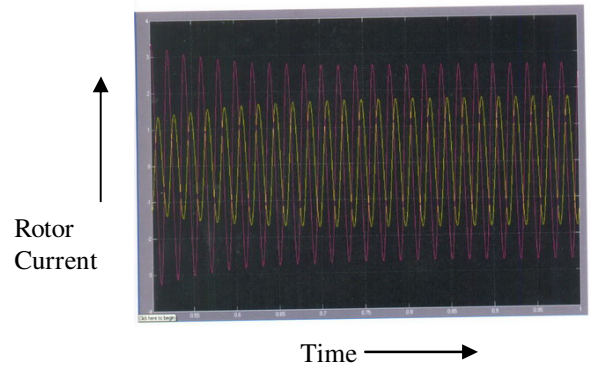

Fig. 6 Rotor Current of Unsymmetrical 2-phase Induction Motor

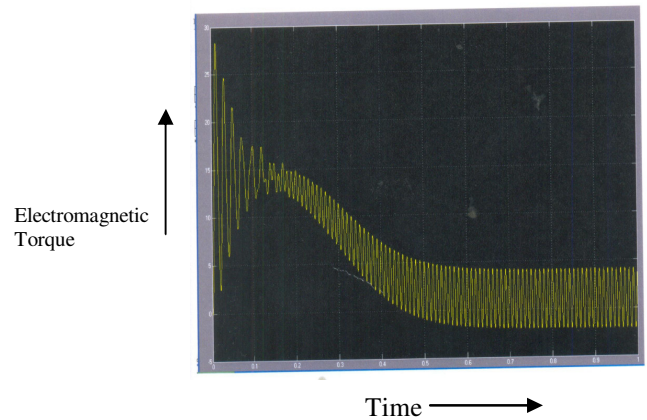

Fig. 7 Electromagnetic Torque of Unsymmetrical 2-phase Induction Motor

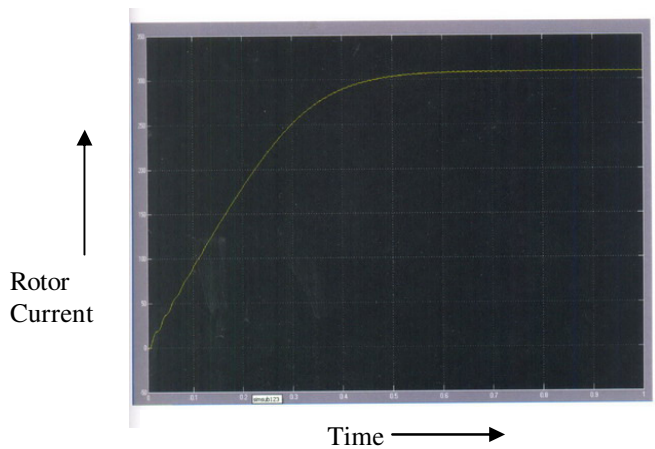

Fig. 8 Rotor Speed of Unsymmetrical 2-phase Induction Motor

The simulation result for unsymmetrical 2-phase induction motor which consists in using dynamic modeling is shown in following figures. The main and auxiliary supply voltages are shown in fig. 4. The main and auxiliary supply voltages are identical in amplitude at 50 Hz. The simulation results of the dynamic response are shown in fig. 5 and in fig. 6. The fig. 5 shows the main and auxiliary stator currents and the fig. 6 shows the main and auxiliary rotor currents. The fig. 7 and fig.8 shows the electromagnetic torque and rotor speed respectively. As the results, in fig. 7 it is seen that the electromagnetic torque response is generated in transient and steady-state.

## 4. Conclusion and Future Scope

The equations which describe the transient and steady state behavior of symmetrical 2-phase induction motor can be set forth, and the computer representation can be developed from them. The equivalent circuits and the computer representation of the symmetrical 2-phase induction motor are quite general and can be readily modified to study a variety of applications of this type of motor. Although the less common modes of operations can also be investigated. For example the equivalent circuits and the computer representation of the unsymmetrical 2-phase induction motor are valid, regardless of the form or the time relationship of the stator applied voltage, generator action can be studied.

The use of computer as an instructional aid to demonstrate the dynamic performance of the various types of the electric motors is of academic importance. The representation which will be developed for the single phase induction motor enable demonstration of the complete dynamic performance of an important and widely used type of electric motor.

## Acknowledgement

Authors are deeply indebted to Priyadarshini Institute of Engineering and Technology, Nagpur for their constant support

## References

[1] Ashfaq Husain, 'Electrical Machines', Second edition, Dhanpat Rai and Co., Educational and Technical Publishers, Delhi 2005.
[2] Gopal K. Dubey, 'Fundamentals of electrical drives', second edition, Narosa publishing house, Delhi 2001.
[3] P. C. Krause, O. Wasynczuk, S. D. Sudhoff, 'Analysis of electric machinery and drive system', second edition, IEEE Press.
[4] P. C. Krause, 'Simulation of unsymmetrical 2-phase induction motor', IEEE transactions on power apparatus and systems, Vol. PAS-84, No. 11, page no. 1025-1037.
[5] Yuttana Kumsuwan, Watcharin Srirattanawichaikul and Suttichai Premrudeepreechacharn, 'Analysis of a 2-phase Induction motor using dynamic model based on MATLAB/Simulink', International conference on the role of universities in hands-on education Rajamangala university of technology lanna, Chiang-Mai, Thailand, 23-29 August 2009.